\documentclass[twocolumn]{aastex701}

\newcommand{\target}{KSP-OT-202104a}

\newcommand{\hi}{\hbox{H\,{\sc i}}}

\newcommand{\hei}{\hbox{He\,{\sc i}}}
\newcommand{\heii}{\hbox{He\,{\sc ii}}}

\def\arcmin{\hbox{$^\prime$}}
\def\arcsec{\hbox{$^{\prime\prime}$}}
\def\simgt{\lower.5ex\hbox{$\; \buildrel > \over \sim \;$}}%
\def\simlt{\lower.5ex\hbox{$\; \buildrel < \over \sim \;$}}%
\def\vi{\mbox{$V\!-\!I$}}%
\def\bi{\mbox{$B\!-\!I$}}%

\begin{document}

\title{A New WZ Sagittae-type Dwarf Nova {\target} Near the Period Minimum from the KMTNet Supernova Program}  
\shorttitle{S. C. Kim et al.}

\correspondingauthor{Youngdae Lee}
\email{hippo206@gmail.com}

\author[0000-0001-9670-1546]{Sang Chul Kim}
\affil{Korea Astronomy and Space Science Institute 776, Daedeokdae-ro, Yuseong-gu, Daejeon 34055, Republic of Korea}
\affil{Korea University of Science and Technology (UST), 217 Gajeong-ro, Yuseong-gu, Daejeon 34113, Republic of Korea}
\email{sckim@kasi.re.kr}

\author[0000-0002-6261-1531]{Youngdae Lee}
\affil{Korea Astronomy and Space Science Institute 776, Daedeokdae-ro, Yuseong-gu, Daejeon 34055, Republic of Korea}
\affil{Department of Astronomy and Space Science, Chungnam National University, Daejeon 34134, Republic of Korea}
\affil{Research Institute of Natural Sciences, Chungnam National University, Daejeon 34134, Republic of Korea}
\email{hippo206@gmail.com}

\author[0000-0003-4200-5064]{Dae-Sik Moon} 
\affil{David A. Dunlap Department of Astronomy and Astrophysics, University of Toronto, 50 St. George Street, Toronto, ON M5S 3H4, Canada}
\email{moon@astro.utoronto.ca}

\author[0000-0002-3505-3036]{Hong Soo Park} 
\affil{Korea Astronomy and Space Science Institute 776, Daedeokdae-ro, Yuseong-gu, Daejeon 34055, Republic of Korea}
\affil{Korea University of Science and Technology (UST), 217 Gajeong-ro, Yuseong-gu, Daejeon 34113, Republic of Korea}
\email{hspark@kasi.re.kr}

\author[0000-0003-3656-5268]{Yuan Qi Ni}
\affil{Kavli Institute for Theoretical Physics, University of California, Santa Barbara, 552 University Road, Goleta, CA 93106-4030, USA}
\affil{Las Cumbres Observatory, 6740 Cortona Drive, Suite 102, Goleta, CA 93117, USA}
\affil{David A. Dunlap Department of Astronomy and Astrophysics, University of Toronto, 50 St. George Street, Toronto, ON M5S 3H4, Canada}
\email{chrisni@kitp.ucsb.edu}

\author[0000-0001-8748-0016]{Nan Jiang}
\affil{David A. Dunlap Department of Astronomy and Astrophysics, University of Toronto, 50 St. George Street, Toronto, ON M5S 3H4, Canada}
\email{steven.jiang@mail.utoronto.ca}

\author[0000-0002-5742-8476]{Hyobin Im}
\affil{Korea Astronomy and Space Science Institute 776, Daedeokdae-ro, Yuseong-gu, Daejeon 34055, Republic of Korea}
\affil{Korea University of Science and Technology (UST), 217 Gajeong-ro, Yuseong-gu, Daejeon 34113, Republic of Korea}
\email{hyobin@kasi.re.kr}

\shorttitle{WZ Sge-type Dwarf nova : {\target}}
\shortauthors{S. C. Kim et al.}

\begin{abstract}
We present photometric and spectroscopic studies of a new WZ Sagittae (Sge)-type dwarf nova (DN) 
  {\target} discovered by the Korea Microlensing Telescope Network 
  Supernova Program.
The source exhibits outburst amplitudes of $\sim 8$ mag with a duration of $\sim 28.5$ days in the $V$-band.
It is a type D DN among WZ Sge-types, and we estimate the superhump period
  to be $P_{\rm sh} \approx 71.7$ minutes ($=0.04978$ days).
Its spectrum shows blue continuum 
  as often found in optically-thick accretion disks of DNe during outbursts
  with hydrogen absorption lines from H$\beta$ to H$\zeta$.
Since the orbital period in WZ Sge-type DNe is typically very close to the superhump period,
  we consider that this target would belong to the small sample of DNe
  below the period minimum and may be evolving toward AM Canum Venaticorum (AM CVn) stars.
This system therefore adds an example of a short-period dwarf nova
  with a low mass-transfer rate to the known sample.
\end{abstract}


\section{Introduction} \label{sec:intro}

Dwarf novae (DNe) are binary stellar systems composed of a white dwarf (WD) primary and
  a lower-mass secondary companion, forming
  an important population of cataclysmic variables (CVs).
The Roche lobe-filling secondary transfers mass to the primary  through 
the inner Lagrangian point to form an accretion disk that leads to 
thermal instability-induced outbursts with amplitudes mostly in
the range of 2--10 mag  
and durations of days to weeks \citep{Can87, warner95, osaki96, Patterson96, Drake14}.
DNe have been classified into three groups based on their outburst shapes:
  U Gem (Geminorum)-type with repetition of similar outbursts,
  SU UMa (Ursae Majoris)-type showing superoutbursts and normal outbursts, and
  Z Cam (Camelopardalis)-type displaying standstills with intermediate brightness
    between the outburst peak and quiescent phase brightness \citep{warner95, Otula16}.
SU UMa-type DNe are further classified into three sub-classes
  by mass transfer rate to
  ER UMa-, SU UMa-, and WZ Sge-types in the decreasing order of mass transfer rate \citep{Otula16}.
ER UMa-type DNe have relatively short time intervals between the outbursts \citep{lee24},
  while infrequent outbursts ($\sim$ once a decade) of WZ Sge-type DNe
  show large outburst brightness changes of $\sim$ 6--10 mag 
  \citep{odonoghue91,Patterson93,Osaki95,KatoEtal01, Kato15WZSge, kato17,Tampo23, Tampo25, 
  Tampo2512, Tampo26}.

SU UMa-type DNe show periodic magnitude oscillations of 
  up to $\sim 0.2$ mag during superoutbursts, which are called ``superhumps''
  considered to be from the precessing, tidally distorted accretion disk
  \citep{Whitehurst88, Hirose90, kato09Dec, Kato15WZSge}.
The periods of the superhumps ($P_{\rm sh}$) are typically
  a few percent longer than orbital periods ($P_{\rm orb}$).
The superoutburst light curves (LCs) divide into three stages of A, B, and C \citep{kato09Dec}:
  `A' for an initial evolutionary phase with a longer superhump period,
  `B' for intermediate stage with systematically varying periods, and
  `C' for a short and almost constant period.
The superhumps at stage A are considered to be driven by the dynamical precession
  of the 3:1 resonance in the accretion disk \citep{Osaki13b}.

Compared to ordinary SU UMa-types, the mass transfer rates of WZ Sge-types are  
  smaller than $\simlt 10^{15}$ g s$^{-1}$ ($1.6 \times 10^{-10} M_{\odot}$ yr$^{-1}$;
  \citealt{howell95,osaki02,Killestein25}), 
  e.g., even $\sim 10^{13}$ g s$^{-1}$ 
  ($10^{-13} M_{\odot}$ yr$^{-1}$; \citealt{Neustroev23})
  with much smaller viscosity during quiescence.
This low viscosities of WZ Sge-types may be more compatible with 
brown dwarf (BD) secondaries without magnetic activity and 
with small mass ratios of $q = M_2 / M_1$ $\simlt$ 0.1 
where $M_2$ and $M_1$ are the masses of the secondary and the primary, respectively 
\citep{osaki02}. 

As the mass transfer from the low mass secondary to the primary in a DN continues,
the angular momentum, orbital period, and orbital separation decrease 
to reach the ``period minimum'' of about 76 minutes 
\citep{PaczynskiSienkiewicz81,Kolb99,Thorstensen02,Gansicke09,
  knigge11,McAllister19,MunozGiraldo24,Krushevska24}.
(Note that a broad range for the period minimum of $65-80$ minutes have been adopted
in literature.)
The mass loss and developed degeneracy in the secondary induce it to
become a BD-like object with mass $\simlt$ 0.07 $M_\odot$ 
and to terminate its contraction \citep{Neustroev17, kato17}.
The orbital period after this begins to increase to conserve angular momentum,
and the DN become ``period bouncers'' as evolving back towards a longer orbital period
  \citep{King88,Pala18,MunozGiraldo24,Killestein25}.
Consequently, DNe are believed to evolve initially with a main sequence secondary
as its orbital period decreases and later to evolve 
with a degenerate secondary as the orbital period increases \citep{uemura02} 

Of particularly interest are those DNe with periods smaller than the period minium. 
Only handful of these objects have been identified \citep[e.g.,][]{Green20,lee22}, 
  and their evolutionary paths remain uncertain. 
Although at least part of them are thought to evolve eventually to 
  AM Canum Venaticorum (AM CVn) stars \citep[ultracompact binaries
  with a WD primary and a He star or He WD secondary having
  the most shortest orbital period of
  $<70$ minutes;][]{Podsiadlowski03, Nelemans05, Solheim10, KatoKojiguchi21, Aungwerojwit25, Kojiguchi26},
  whether or not some of them evolve back to longer periods is poorly understood.
Also, whether their evolutions depend on their DN types remains unknown.
In our previous studies \citep{lee22,lee24}, we report discoveries of 
SU UMa-like (KSP-OT-201701a) and ER UMa-type (KSP-OT-201712a) DNe with large mass 
transfers below the period minimum that show transitional nature with small He abundance.
Here, we present the discovery and  photometric 
and spectroscopic studies of a new  WZ Sge-type DN {\target} 
under the period minium discovered in the Korea Microlensing Telescope Network (KMTNet)
Supernova Program \citep{Moon16}.

This paper is organized as follows. 
We describe our observations and data reduction in \S2
  followed by photometric analysis of LCs,
  period analysis, color evolution, and spectroscopic results
  in \S3.
In \S4, we discuss the DNe with periods shorter than the period minimum
  and their evolution.
We summarize our results in \S5.
    
\section{Observations and Discovery}\label{sec:discovery}  

\target\ was discovered on 2021 April 9 (UT) by the KMTNet \citep{Kim16}
Supernova Program (KSP; \citealt{Moon16})
using three wide-field 1.6-m telescopes in Chile, South Africa, and Australia. 
KSP focuses on discovering early optical transients such as
infant supernovae \citep[e.g.,][]{Afsari19,Moon21,Ni22,Ni23_946,Ni23_959,Ni25,Jiang25,Chang26} 
at a high cadence in the Johnson $BVI$-bands. 
The program also provides opportunities of serendipitously 
discovering other types of transients such as DNe  \citep[e.g.,][]{brown18,lee19,lee22,lee24}.
The discovery of \target\ was made by the Australian telescope of KMTNet 
  at magnitudes
  $B  = 19.28 \pm 0.03$ (UT 09:33), 
  $V = 18.81 \pm 0.05$ (UT 09:35), and 
  $I = 18.57 \pm 0.05$ (UT 09:37)\footnote{These magnitudes are standardized values
  after applying $B$-band color correction and $I$-band magnitude shift
  which are explained in the next paragraph, but before the extinction correction.}
at the location of ($\alpha, \delta$) =
($\rm 08^h14^m20.14^s$, $\rm -26\degr02\arcmin37\farcs21$) 
= (123.58390\degr, --26.04367\degr) (J2000)\footnote{$(l,b)$ = (245.45\degr, +4.75\degr)}, and it was subsequently detected by the other two telescopes.
The source was not detected on images obtained just previously by the Chilean telescope
about 8 hours earlier than the first detections above the $\sim$ 21.5 limiting detection magnitudes of the image.

Point-spread function photometry was performed  
using the DAOPHOT-II package \citep{Ste87} on the KSP images, 
and the photometric standardization was made by comparing 
the brightness of \target\ with those of multiple nearby standard
stars from the AAVSO Photometric All Sky Survey (APASS) database
on the same images.\footnote{https://www.aavso.org/apass} 
This standardization method of KMTNet images requires a $B$-band color correction of
$\Delta$$m_B$ = $m_B - m_{B'}$ $\simeq$ 0.27 $(m_B-m_V)$ + \textrm{offset},
where $m_B$ and $m_{B'}$ are color-corrected $B$-band magnitudes 
in the standard Johnson system 
and uncorrected KTMNet $B$-band magnitudes, respectively \citep{Par17,Ni25}.
Since the AAVSO photometry is available with the $BVi$-bands,
  we applied another correction of $I$ = $i$ -- 0.4 mag \citep[see][]{Par17}.
The reddening is estimated to be $E(B-V) \le 0.05$ mag
  assuming the distance $\le 2$ kpc \citep{Green18JOSS, Green19}.
Table~{\ref{tab_app}} shows the photometric catalog for {\target}.

\begin{table*} [h!]
\centering
\caption{Photometric catalog for {\target}} \label{tab_app}
\renewcommand{\arraystretch}{0.6}
\begin{tabular}{c c c | c c c | c c c}

\hline
\hline
\multicolumn{3}{c|}{$B$} & \multicolumn{3}{c|}{$V$}  & \multicolumn{3}{c}{$I$}           \\
  MJD       &  Mag  &  Error     &    MJD        &   Mag &  Error     &     MJD       & Mag   &  Error      \\
$[$days$]$  &  $[$mag$]$  &  $[$mag$]$       &    $[$days$]$       &   $[$mag$]$ &  $[$mag$]$       &     $[$days$]$      & $[$mag$]$   &  $[$mag$]$      \\
\hline
59313.3984 & 19.28 & 0.03 & 59313.3999 & 18.81 & 0.05 & 59313.4013 & 18.57 & 0.05 \\
59313.4564 & 18.55 & 0.01 & 59313.4579 & 18.08 & 0.02 & 59313.4594 & 18.09 & 0.03 \\
59313.5138 & 18.05 & 0.01 & 59313.5153 & 17.57 & 0.01 & 59313.5558 & 17.35 & 0.02 \\
59313.5529 & 17.76 & 0.01 & 59313.5544 & 17.31 & 0.01 & 59313.7499 & 16.36 & 0.01 \\
59313.7470 & 16.65 & 0.00 & 59313.7485 & 16.27 & 0.01 & 59313.7969 & 16.27 & 0.01 \\
59313.7938 & 16.51 & 0.00 & 59313.7954 & 16.14 & 0.01 & 59314.0117 & 16.08 & 0.01 \\
59314.0088 & 16.36 & 0.00 & 59314.0102 & 15.99 & 0.00 & 59314.0632 & 16.13 & 0.01 \\
59314.0604 & 16.40 & 0.00 & 59314.0618 & 16.03 & 0.01 & 59314.3947 & 16.23 & 0.01 \\
59314.3917 & 16.50 & 0.01 & 59314.3932 & 16.10 & 0.01 & 59314.4439 & 16.28 & 0.02 \\
59314.4408 & 16.58 & 0.00 & 59314.4424 & 16.18 & 0.01 & 59314.7480 & 16.31 & 0.01 \\
59314.7450 & 16.61 & 0.01 & 59314.7465 & 16.23 & 0.01 & 59314.8005 & 16.36 & 0.01 \\
59314.7977 & 16.68 & 0.00 & 59314.7990 & 16.28 & 0.01 & 59315.0073 & 16.38 & 0.01 \\
59315.0044 & 16.71 & 0.00 & 59315.0058 & 16.32 & 0.00 & 59315.0573 & 16.40 & 0.01 \\
59315.0545 & 16.72 & 0.00 & 59315.0559 & 16.33 & 0.00 & 59315.4515 & 16.47 & 0.02 \\
59315.4486 & 16.84 & 0.01 & 59315.4501 & 16.40 & 0.01 & 59315.7482 & 16.57 & 0.01 \\
59315.7454 & 16.88 & 0.01 & 59315.7467 & 16.49 & 0.01 & 59315.7968 & 16.58 & 0.01 \\
59315.7938 & 16.91 & 0.01 & 59315.7953 & 16.52 & 0.01 & 59317.0099 & 16.81 & 0.01 \\
59317.0070 & 17.19 & 0.01 & 59317.0084 & 16.76 & 0.01 & 59317.0716 & 16.80 & 0.01 \\
59317.0689 & 17.18 & 0.01 & 59317.0702 & 16.78 & 0.01 & 59317.3962 & 16.94 & 0.02 \\
59317.3932 & 17.32 & 0.01 & 59317.3947 & 16.86 & 0.01 & 59317.4483 & 16.94 & 0.01 \\
59317.4454 & 17.30 & 0.01 & 59317.4468 & 16.88 & 0.01 & 59317.5038 & 16.93 & 0.01 \\
59317.5011 & 17.31 & 0.01 & 59317.5024 & 16.87 & 0.01 & 59317.5418 & 16.92 & 0.01 \\
59317.5389 & 17.32 & 0.01 & 59317.5404 & 16.89 & 0.01 & 59317.7456 & 16.97 & 0.02 \\
59317.7426 & 17.32 & 0.01 & 59317.7441 & 16.89 & 0.01 & 59317.7932 & 16.95 & 0.01 \\
59317.7902 & 17.35 & 0.01 & 59317.7917 & 16.95 & 0.01 & 59321.0036 & 17.45 & 0.02 \\
59321.0009 & 17.85 & 0.01 & 59321.0023 & 17.43 & 0.01 & 59321.0506 & 17.48 & 0.03 \\
59321.0479 & 17.87 & 0.01 & 59321.0493 & 17.47 & 0.02 & 59321.3907 & 17.49 & 0.02 \\
59321.3878 & 17.90 & 0.01 & 59321.3892 & 17.45 & 0.01 & 59321.4373 & 17.46 & 0.02 \\
59321.4345 & 17.93 & 0.01 & 59321.4360 & 17.44 & 0.01 & 59321.7447 & 17.49 & 0.02 \\
59321.7417 & 17.91 & 0.01 & 59321.7432 & 17.48 & 0.01 & 59321.7964 & 17.48 & 0.02 \\
59321.7934 & 17.92 & 0.01 & 59323.0066 & 17.63 & 0.01 & 59323.0080 & 17.57 & 0.02 \\
59323.0052 & 18.09 & 0.01 & 59323.0542 & 17.64 & 0.01 & 59323.0556 & 17.59 & 0.02 \\
59323.0527 & 18.12 & 0.01 & 59323.3934 & 17.67 & 0.02 & 59323.3948 & 17.72 & 0.03 \\
59323.3919 & 18.15 & 0.02 & 59323.4402 & 17.67 & 0.02 & 59323.4416 & 17.64 & 0.03 \\
59323.4387 & 18.17 & 0.02 & 59323.4873 & 17.65 & 0.02 & 59323.4887 & 17.65 & 0.02 \\
59323.4859 & 18.16 & 0.02 & 59323.5243 & 17.66 & 0.01 & 59323.5258 & 17.64 & 0.02 \\
59323.5228 & 18.14 & 0.01 & 59325.0039 & 17.81 & 0.02 & 59325.0080 & 17.75 & 0.02 \\
59324.9989 & 18.37 & 0.02 & 59325.0066 & 17.80 & 0.02 & 59325.0575 & 17.82 & 0.03 \\
59325.0053 & 18.33 & 0.02 & 59325.0560 & 17.82 & 0.02 & 59325.3877 & 18.00 & 0.04 \\
59325.0546 & 18.35 & 0.02 & 59325.3863 & 17.94 & 0.03 & 59325.4362 & 17.92 & 0.04 \\
59325.3847 & 18.46 & 0.03 & 59325.4347 & 17.91 & 0.03 & 59325.4828 & 17.98 & 0.05 \\
59325.4332 & 18.40 & 0.03 & 59325.4813 & 17.90 & 0.03 & 59325.5253 & 17.88 & 0.04 \\
59325.4798 & 18.37 & 0.03 & 59325.5239 & 17.76 & 0.03 & 59325.8217 & 17.82 & 0.03 \\
59325.5224 & 18.23 & 0.03 & 59325.8202 & 17.89 & 0.02 & 59328.9921 & 18.07 & 0.03 \\
59325.8189 & 18.30 & 0.02 & 59328.9908 & 18.11 & 0.02 & 59329.0448 & 18.09 & 0.02 \\
59328.9894 & 18.60 & 0.02 & 59329.0434 & 18.11 & 0.02 & 59329.3873 & 18.14 & 0.04 \\
59329.0420 & 18.63 & 0.02 & 59329.3858 & 18.20 & 0.04 & 59330.9959 & 18.25 & 0.03 \\
59329.3844 & 18.68 & 0.05 & 59329.4323 & 18.27 & 0.05 & 59331.0533 & 18.29 & 0.04 \\
59330.9930 & 18.89 & 0.03 & 59330.9944 & 18.36 & 0.03 & 59331.4888 & 18.33 & 0.05 \\
59331.0019 & 18.86 & 0.03 & 59331.0520 & 18.42 & 0.04 & 59331.7315 & 18.30 & 0.03 \\
59331.0505 & 18.92 & 0.04 & 59331.7300 & 18.37 & 0.03 & 59331.7793 & 18.31 & 0.03 \\
59331.3861 & 18.88 & 0.05 & 59331.7778 & 18.38 & 0.03 & 59332.9896 & 18.35 & 0.03 \\
59331.7285 & 18.90 & 0.03 & 59332.9882 & 18.46 & 0.02 & 59333.0398 & 18.36 & 0.03 \\
59331.7763 & 18.91 & 0.03 & 59333.0385 & 18.51 & 0.03 & 59333.3862 & 18.44 & 0.04 \\
59332.9868 & 18.96 & 0.01 & 59333.3848 & 18.50 & 0.03 & 59333.4333 & 18.54 & 0.05 \\
59333.0370 & 18.98 & 0.03 & 59333.4318 & 18.59 & 0.04 & 59333.4802 & 18.52 & 0.05 \\
59333.3833 & 19.03 & 0.02 & 59333.4788 & 18.53 & 0.04 & 59334.9930 & 18.52 & 0.04 \\
59333.4305 & 19.08 & 0.04 & 59333.7472 & 18.54 & 0.03 & 59335.0428 & 18.55 & 0.04 \\
59333.7459 & 19.06 & 0.03 & 59334.9916 & 18.65 & 0.02 & 59335.7800 & 18.61 & 0.03 \\
59334.9903 & 19.14 & 0.02 & 59335.0415 & 18.69 & 0.03 & 59336.0137 & 18.58 & 0.04 \\
59335.0400 & 19.17 & 0.02 & 59335.7785 & 18.62 & 0.02 & 59336.0667 & 18.63 & 0.04 \\
59335.7768 & 19.19 & 0.02 & 59336.0123 & 18.73 & 0.03 & 59336.9899 & 18.67 & 0.04 \\
59336.0108 & 19.18 & 0.02 & 59336.0653 & 18.72 & 0.03 & 59336.9921 & 18.73 & 0.04 \\
59336.0639 & 19.25 & 0.02 & 59336.0682 & 18.71 & 0.04 & 59337.0452 & 18.66 & 0.04 \\
59336.1079 & 19.28 & 0.03 & 59336.1093 & 18.78 & 0.03 & 59338.9946 & 20.28 & 0.21 \\
59336.9870 & 19.32 & 0.02 & 59336.9885 & 18.75 & 0.02 & 59339.0493 & 20.41 & 0.19 \\
59337.0425 & 19.31 & 0.02 & 59337.0439 & 18.79 & 0.03 & 59339.0976 & 20.78 & 0.26 \\
59337.0888 & 19.32 & 0.03 & 59337.0902 & 18.83 & 0.04 & 59339.1007 & 20.81 & 0.25 \\
59337.7275 & 19.41 & 0.02 & 59337.7290 & 18.86 & 0.02 &  -- & -- & --  \\
59337.7770 & 19.41 & 0.02 & 59337.7785 & 18.91 & 0.03 &  -- & -- & --  \\
59338.9919 & 20.71 & 0.37 & 59338.9933 & 20.14 & 0.46 &  -- & -- & --  \\
59339.0464 & 21.33 & 0.10 & 59339.0480 & 20.65 & 0.12 &  -- & -- & --  \\
59339.0949 & 21.33 & 0.12 & 59339.0963 & 20.65 & 0.13 &  -- & -- & --  \\
\hline
\multicolumn{9}{l}{NOTE : Extiction is not corrected.} \\

\end{tabular}
\end{table*}

Spectroscopic observations of \target\ were conducted using 
the Gemini Multi-Object Spectrograph (GMOS) on the 8-m Gemini-South telescope
on 2021 April 23 (UT) (Program ID: GS-2021A-Q-117). 
Four 60-s exposures were obtained for each of
  the blue (3960--7050 \AA\ with B600 grating) and
  red (5405--10000 \AA\ with R400 grating) channels
  with 1\farcs0 slit, resulting in  spectral resolving powers 
  of about 850 (blue) and 950 (red).
The Gemini IRAF\footnote{IRAF is distributed by National Optical Astronomy 
  Observatories, which is operated by the Association of Universities for Research in Astronomy, Inc. (AURA), under cooperative agreement with the National Science Foundation, USA.} package was used to perform basic reduction  of the observed spectra, 
  such as image pre-processing, wavelength calibration,
  and flux calibration. 
Spectro-photometric calibration was obtained by observing 
the standard star GD 108. 

\section{Light Curve Analysis} \label{sec:lightcurve}  
\subsection{Light Curve Shape and Parameters}  \label{sec:LCshape}  

\begin{figure*}   
\epsscale{0.8} 
\plotone{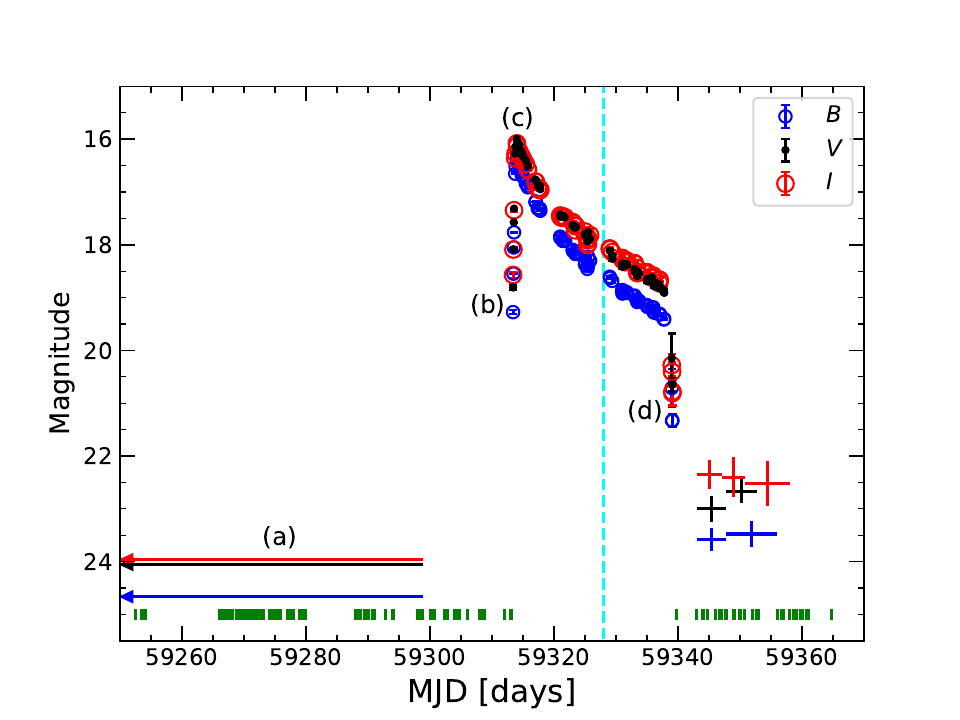}
\caption{$BVI$-band light curves of the superoutburst of {\target} 
  with no extinction correction.
The cyan dashed vertical line marks the epoch when the Gemini spectrum was taken,
and the short green vertical lines at the bottom 
correspond to the epochs outside the outburst when our $V$-band observations were made.
Labels (a)--(d) denote different phases:
(a) quiescent phase prior to the superoutburst in MJD = 58795.27--59298.86;
(b) epoch for the first detection at MJD = 59313.40;
(c) epoch near the peak at MJD = 59314.01; and 
(d) epoch of the last detection at MJD = 59339.05.
The labels (a)--(d) coincide with those in the images shown in Figure~\ref{fig:images}.
The horizontal bars with arrows represent the quiescent magnitudes of the source:
  $B = 24.67 \pm 0.22$, $V = 24.05 \pm 0.21$, and $I = 23.96 \pm 0.36$ mag.
The seven plus symbols near the right bottom corner are brightnesses of the source
in MJD = 59342--59359 during its tail part obtained by stacking images (see text for details).
}
\label{fig:fulllc}
\end{figure*}

Figure~\ref{fig:fulllc} shows the LC of \target\ that we obtained
over a period of $\sim 120$ days, showing an outburst activity
lasting $\sim$ 30 days (MJD 59310--59340) in the middle with 
a peak brighter than quiescence by about eight magnitudes.
The highly asymmetric shape of the outburst featured with a concave
plateau is typical of a WZ Sge-type DN \citep{Kato15WZSge}.
The concave shape is considered to be caused by the viscous depletion 
of a large amount of matter accumulated between consecutive outbursts \citep{osaki96}.

\begin{figure*}   
\epsscale{0.8} 
\plotone{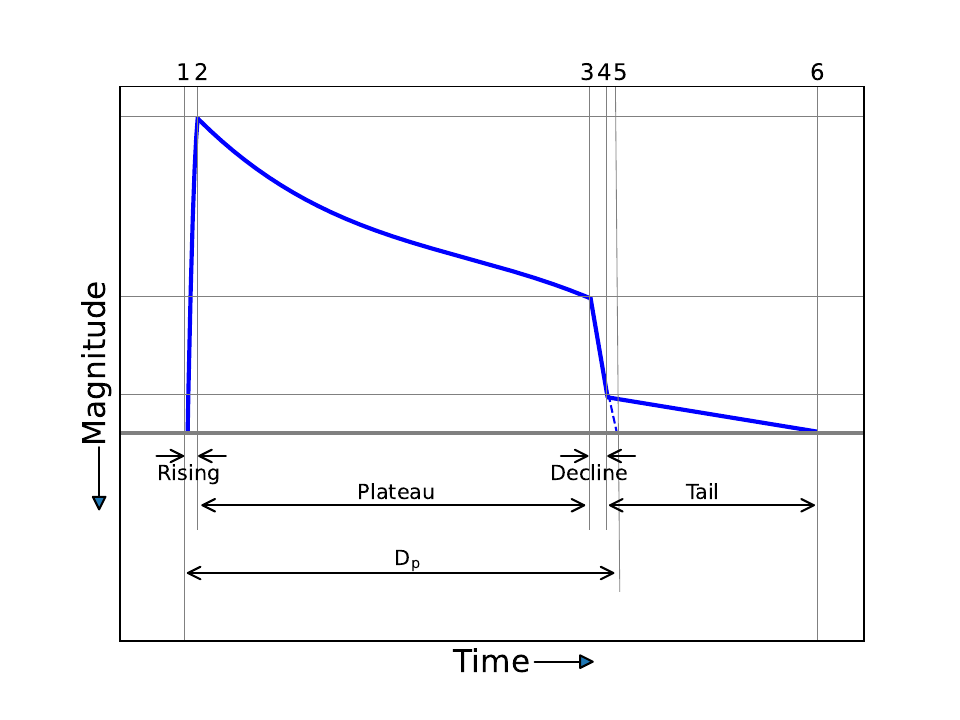}
\caption{Schematic light curve diagram (blue solid curve) 
of a type D WZ Sge-type DN
without rebrightening.
Thick grey horizontal line represents the quiescent phase brightness. 
The beginning points of the four phases---rising, plateau, decline,
and tail parts---are marked as 1, 2, 3, and 4, respectively.
The short blue dashed line is the extension of the decline part
to the quiescent magnitude. 
`D$_{\mathrm{p}}$' between points 1 and 5 is 
the outburst duration above the quiescent magnitude.
}
\label{fig:schematic}
\end{figure*}

\begin{figure*}   
\epsscale{0.8} 
\plotone{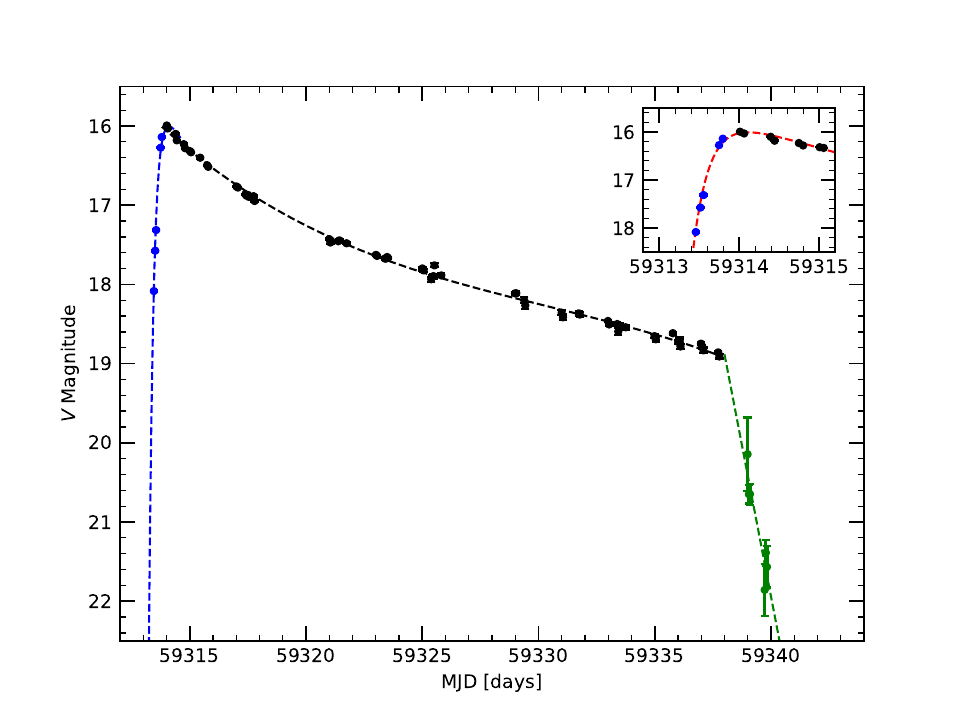}
\caption{Comparison between the observed $V$-band light curve of \target\
  (without extinction correction)
  with the best-fit polynomial fittings:
blue, black, and green colors are
for the observations (filled circles) and fittings (dashed curves)
of the rising, plateau, and decline parts, respectively. 
The inset shows the fitting (red dashed curve) of the near-peak light curve
using the equation
$m_{pe\rm }(d) = A_p(1-e^{-d/r_{\rm pe}})(1+{\alpha}d/d_{\rm pe})$
to obtain the peak magnitude of $V = 16.01$ mag
(see text for details).
}
\label{fig:LC_polynomial}
\end{figure*}

We measure the LC parameters of \target\ using a schematic diagram 
of the typical LCs of type D WZ~Sge-type DNe consisting of 
four phases---rising, plateau, decline, and tail parts---as shown  
in Figure~\ref{fig:schematic} (see below for the classifiction
of \target\ as a type D). 
Figure~\ref{fig:LC_polynomial} shows the results of our fitting
of the $V$-band LC of \target\ by applying separate polynomial fittings 
to the first three phases excluding the tail part.
For the measurements of the peak magnitudes, we additionally fit 
the LCs around the peak area using the equation
$m_{pe\rm }(d) = A_p(1-e^{-d/r_{\rm pe}})(1+{\alpha}d/d_{\rm pe})$,
where $A_p$ is the overall amplitude, 
$d_{\rm pe}$ is a scale length for the rise,
and $\alpha$ is  the plateau slope \citep[see][]{Gio02}.
We obtain 16.01 mag for the $V$-band peak magnitude  
  as well as the outburst amplitude of $\sim 8$ mag 
  and the outburst duration (= D$_{\mathrm{p}}$ in \autoref{fig:schematic})
  of $\sim$ 28.5 days.
Table~\ref{tab:parameters_table} lists the obtained parameters
of the LCs of \target.

Many WZ Sge-type DNe show a rapid decline in brightness
  which is caused by the viscosity decrease in a cold disk \citep{osaki01,Kato15WZSge}
  and show rebrightening, or ``echo outbursts'', 
  later than one day after the rapid decline \citep{osaki02, Patterson02}.
\citet{imada06} classified WZ Sge-type DNe into five types based on rebrightening:
(1) `A' with a long-duration rebrightening;
(2) `B' with multiple discrete rebrightenings;
(3) `C' with a single rebrightening;
(4) `D' with no rebrightening;, and
(5) `E' with double superoutbursts \citep{kato09Dec,Kimura16}. 
They are thought to represent different evolutionary stages of WZ Sge-type DNe
  as C $\rightarrow$ D $\rightarrow$ A $\rightarrow$ B $\rightarrow$ E
  \citep{Kato15WZSge}.  
\target\ appears to be type D given the absence of rebrightening in its LC (see Figure~\ref{fig:fulllc}).

\begin{figure*}   
\epsscale{0.8} 
\plotone{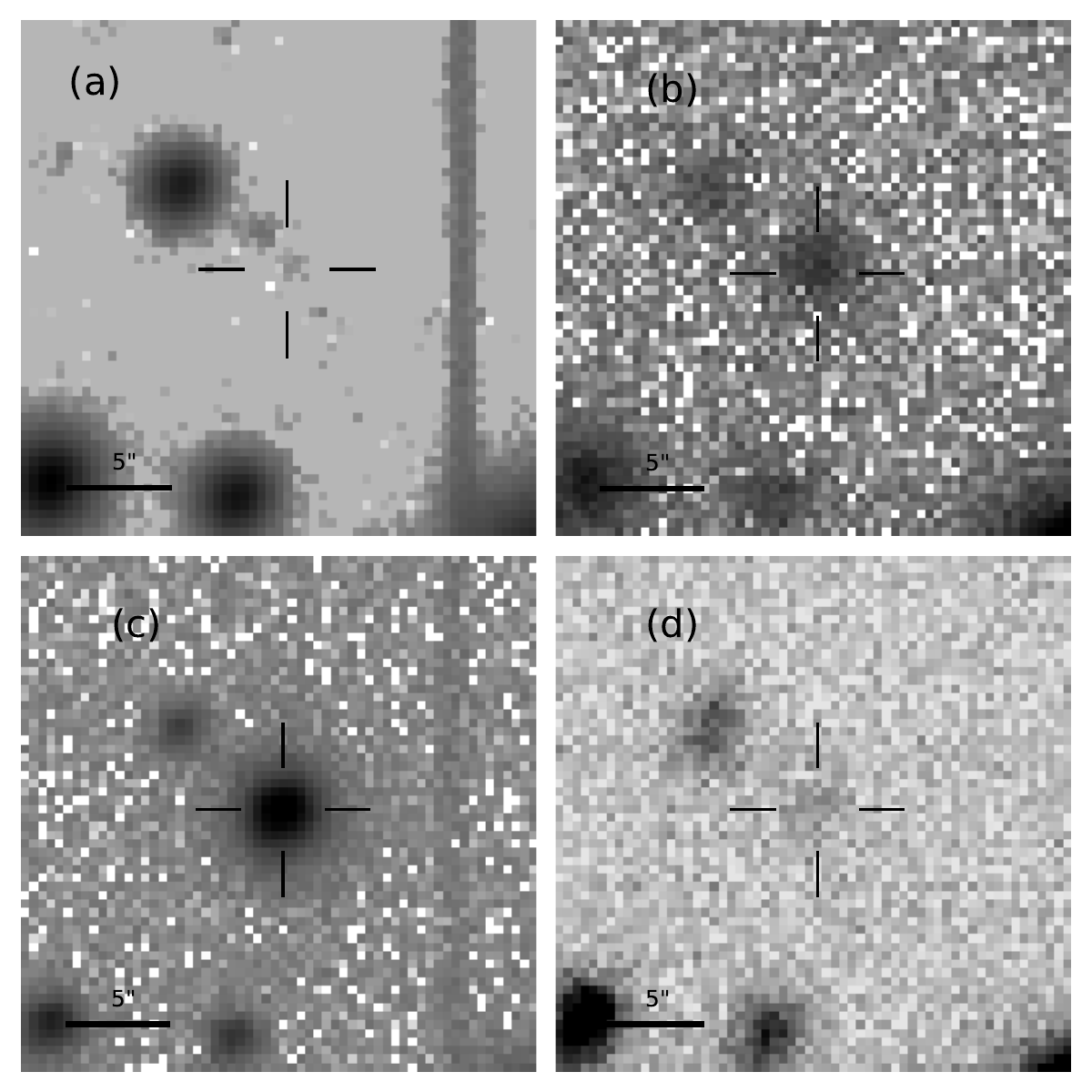}
\caption{$V$-band images of {\target}. 
Panel (a) is a quiescent phase image made by stacking 549 60-s exposures 
  obtained during MJD = 58795.27--59298.86 before the superoutburst.
Panel (b) is the first detection image at MJD = 59313.40;
(c) is a near-peak image of the superoutburst at MJD = 59314.01; and
(d) is the last image during the declining part at MJD = 59339.05.
The labels (a)--(d) coincide with those in the light curve in Figure~\ref{fig:fulllc}.
North is up, and east is to the left. 
Crosshairs mark the position of {\target}.
Horizontal bars of 5\arcsec\ are shown in the bottom-left corners.
}
\label{fig:images}
\end{figure*}

Figure~\ref{fig:images} shows $V$-band images of \target\
from different phases: 
(a) is a deep quiescent image made by stacking 549 60-s exposures 
obtained during MJD = 58795.27--59298.86 which is at least 
14 days ahead of the first detection; 
(b), (c) and (d) are the images of the first detection 
of the rising part at MJD = 59313.40,
near the peak brightness at MJD = 59314.01, and
the last detection of the decline part at MJD = 59339.05, respectively.
We estimate the quiescent phase magnitude of \target\ in \autoref{fig:images} 
  (a) to be $V$ = 24.05 $\pm$ 0.21 mag 
  alongside  its $B$- and $I$-band magnitudes to be 
  $B=24.67 \pm 0.22$ and $I=23.96\pm 0.36$ mag, respectively, 
  using similarly stacked images from the same period as the $V$-band.

After MJD $\simeq$ 59341 when the short decline part appears to end 
(\autoref{fig:fulllc}),
\target\ becomes fainter than the detection limits, which are 
typically in the range of 20--21 mag, of our single
60-s exposures in all the $BVI$-bands.
We, therefore, measure the brightness of \target\ after the decline part
in deep images made by stacking 10 60-s images.
As shown as colored plus symbols in \autoref{fig:fulllc},
we have two deep stacked images for each of the $B$- and $V$-bands
and three images for the $I$-band,
and their mean magnitudes are 
$B\simeq 22.7$, $V \simeq 22.2$, and $I \simeq 22.1$ mag.
We adopt these values as the brightness of the source during
the tail part.
We have not found any sign of rebrightening of \target\
after the plateau part in our images.

\begin{deluxetable*}{c c c c}\label{tab:parameters_table}
		\tablecaption{Outburst Properties of \target}
		\tablehead{
			\colhead{Parameter} & \colhead{$B$} & \colhead{$V$} & \colhead{$I$}
		}
\startdata
Epoch of rising start (in MJD [day]) & $59313.2058\pm0.0041$ & $59313.2024\pm0.0071$ & $59313.1827\pm0.0080$ \\
Rising rate ($\tau_r$ [mag day$^{-1}$]) & $18.91\pm0.29$ & $17.69\pm0.47$ & $16.53\pm0.47$ \\
\hline 
Epoch of peak brightness ($t_p$ in MJD [day]) & $59314.0977\pm0.0030$ & $59314.1124\pm0.0054$ & $59314.1158\pm0.0065$ \\
Peak magnitude([mag]) & $16.38 \pm 0.01$  & $16.01\pm0.01$ & $16.12\pm0.01$ \\
Epoch of plateau end (in MJD [day]) & $59337.8190\pm0.0440$ & $59337.7968\pm0.0554$ & $59337.4052\pm0.1594$ \\
  = Epoch of decline start &&& \\
Magnitude at plateau end & $19.42\pm0.01$ & $18.90\pm0.01$& $18.75\pm0.02$ \\
Decay rate of plateau ($\tau_p$ [mag day$^{-1}$]) & $0.12 \pm 0.01$ & $0.12\pm0.01$ & $0.11\pm0.01$ \\
\hline 
Epoch of decline end (in MJD [day]) & $59341.2653\pm0.2054$ & $59341.6692\pm0.1966$ & $59342.9856\pm0.3774$ \\
  = Epoch of tail start &&& \\  
Decline rate ($\tau_d$ [mag day$^{-1}$]) & $1.53\pm0.10$ & $1.34\pm0.07$ & $0.94\pm0.08$ \\
\hline 
Quiescent phase magnitude [mag] & $24.67\pm0.22$ & $24.05\pm0.21$ & $23.96\pm0.36$ \\ 
Outburst duration (D$_{\mathrm{p}}$ [day]) & $28.06\pm0.21$ & $28.47\pm0.20$ & $29.80\pm0.38$ \\
\enddata
\end{deluxetable*}

\subsection{Period Analysis} \label{sec:Porb}

To obtain the superhump period of \target, 
  we use the phase dispersion minimization \citep[PDM:][]{Stellingwerf78} method,
  which searches for optimum periods from non-sinusoidal pulsation data
  \citep{LinnellNemec85,SchwarzenbergCzerny97,Tampo26}.
Considering the sporadic observations of \target\ inadequate for time-series analysis,
  we also conduct the Lomb-Scargle (LS) analysis
  \citep{lomb76,scargle82}\footnote{Astropy.stats.LombScargle is implemented
  \citep[\url{http://docs.astropy.org/en/stable/stats/lombscargle.html};][]{Ast13,Ast18}}.
To apply these two methods, 10th-degree polynomial fits are subtracted
  from each of the $B$-, $V$-, and $I$-band light curves of the source
  during the plateau part (MJD = 59315–59337), and
  the resulting residual magnitudes from the three bands are combined
  to improve statistical reliability.

We obtain the periods from the PDM and LS methods
  using the midterm period data of the plateau part
  emulating it as stage B among the three evolutionary stages of A, B, and C
  \citep{kato09Dec, Tarasenkov25}
  with 1,000 bootstrap resampling.
Since we are not able to find clear boundaries of stage B from our data,
  we select an optimum range in the middle of the plateau part
  that is as narrow as possible while still including sufficient data.

\begin{figure*}   
\epsscale{1.0}
\plotone{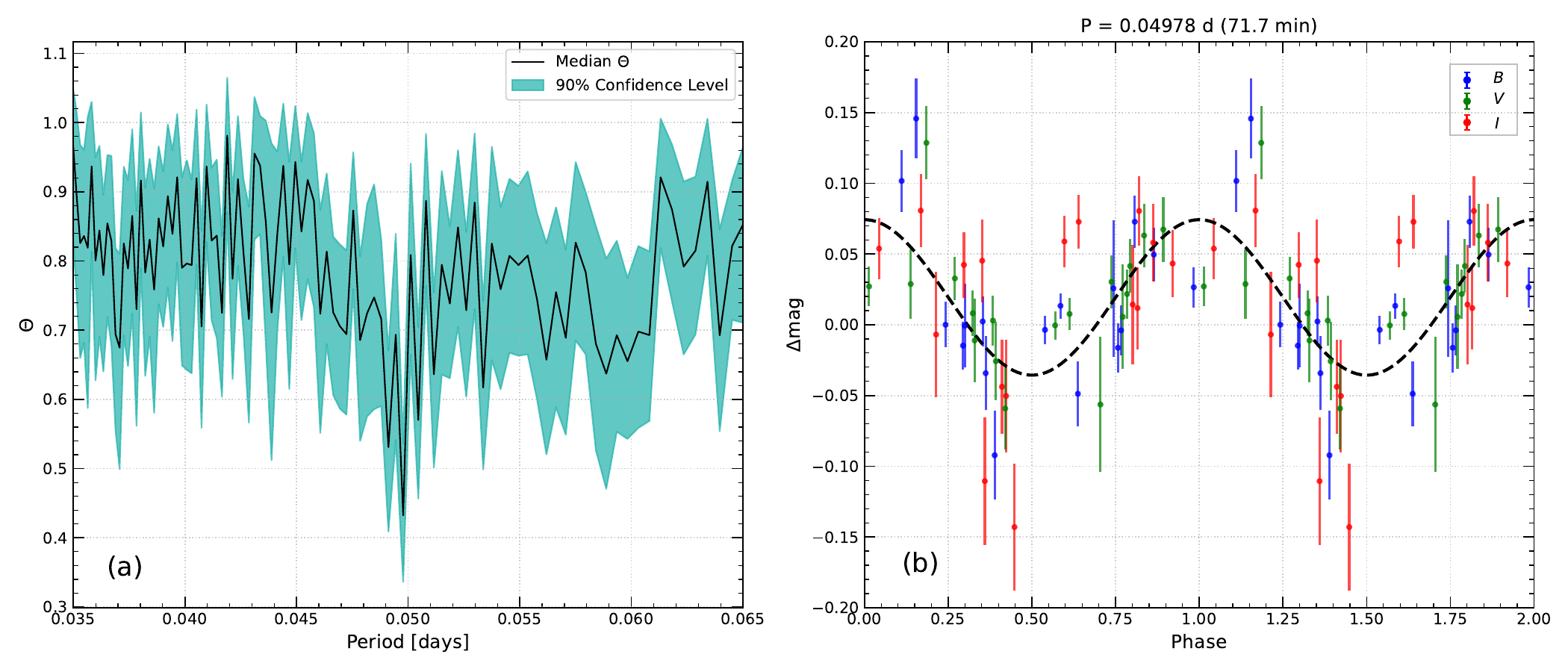}
\caption{(a) Results of PDM analysis and 
  (b) phase-folded profile. 
In panel (a), the black solid line is median $\Theta$ and
  the shaded area presents 90\% confidence range from 1,000 bootstrap resampling.
In panel (b), blue, green, and red colors indicate $B$-, $V$-, and $I$-band, respectively.
The black dashed curve is a sine curve fitted to the data with the period of 71.7 min.
}
\label{fig:PDM}
\end{figure*}

Figure~\ref{fig:PDM} (a) shows the PDM result.
The black line and the shaded area indicate median $\Theta$ and
  90\% confidence range from the 1,000 bootstrap PDM, respectively.
The minimum value of $\Theta$ corresponds to a period of 71.7 minutes ($P_{\rm sh}=0.04978$ days).
Based on this period, phase-folded profile is shown in Figure~\ref{fig:PDM} (b).
The $\Delta$mag of $B$-, $V$-, and $I$-bands are shown as blue, green, and red dots, respectively.
A sine curve fitted to all the data points
  with the period of 71.7 min is presented with black dashed curve.
The LS method also yields the same period of 71.7 min, and
  the consistency between the two independent methods enhances confidence in this period estimate.

\subsection{Color Evolution}\label{sec:LCcolor}

\begin{figure*}   
\epsscale{0.8} 
\plotone{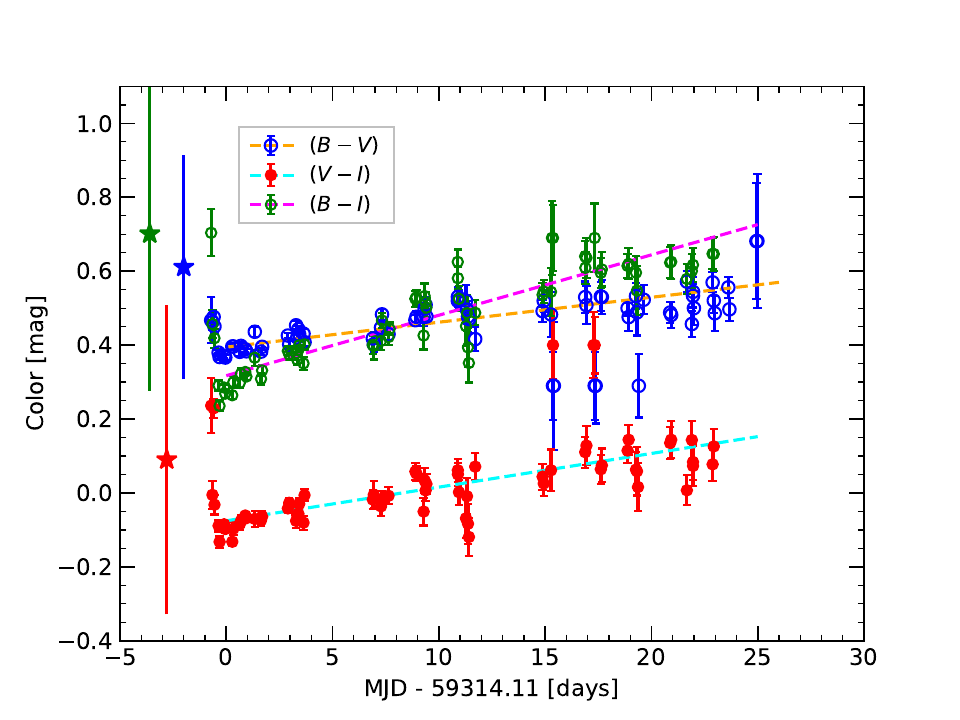}
\caption{Color evolution of \target:
  blue, red, and green circles for (\bv), (\vi), and (\bi) colors
  without extinction correction.
The $x$-axis represents days from the $V$-band peak brightness at MJD = 59314.11.
The stars with large error bars in the left are colors for the quiescent phase
shown for comparison, 
and the dashed lines show the results of linear fittings of the post-peak color evolution.
}
\label{fig:color_evol}
\end{figure*}

Figure~\ref{fig:color_evol} shows the evolution of the
$(B-V)$, $(V-I)$, and $(B-I)$ colors (circles) of \target\ during the outburst.
All the three colors show initial blueward evolution for about one day 
by $\simlt$ 0.4 mag before the peak followed by continuous reddening 
until the end of the outburst  in about 25 days. 
The bluest colors take place near the peak at $(B-V)$ $\approx$ +0.37, 
  $(V-I) \approx -0.13$, and $(B-I) \approx +0.24$ mag.
The post-peak reddening rates of the colors 
  obtained by linear fitting (dashed lines in Figure~\ref{fig:color_evol}) are 
  0.0067, 0.0091, and 0.0163 mag day$^{-1}$ for (\bv),
  (\vi), and  (\bi), respectively, and  
the average colors during the outburst are 
  $(B-V) \simeq 0.46$, $(V-I) \simeq 0.0$, and $(B-I) \simeq 0.43$ mag.
Towards the end of the outburst, the color values of \target\ 
appear to become similar to  
those (stars in Figure~\ref{fig:color_evol}) obtained during the quiescent phase  
prior to the outburst (\S\ref{sec:LCshape}).

The $B-I$ colors of WZ Sge- and ordinary SU UMa-type DNe during outbursts have
been reported to be in the range between --0.13 and 0.25 mag
\citep[e.g.,][]{Neustroev17,lee19,Krushevska24}, 
  and this is consistent with what we obtain in \target\
  if we apply the extinction corrections.
The $B-V$ colors of \target, if extinction correction is applied, 
  also gets closer to the values obtained 
  for a WZ Sge-type DN ASASSN-19oc ($B-V\sim0$) by \citet{Krushevska24} and
  for the 1978 outburst of WZ Sge itself ($B-V = -0.09$) by \citet{Patterson1978}, \citet{Brosch1980},
  and \citet{Howarth81}.
 
\subsection{Spectroscopic Results}\label{sec:SpecSection}

\begin{figure*}   
\epsscale{1.0}
\plotone{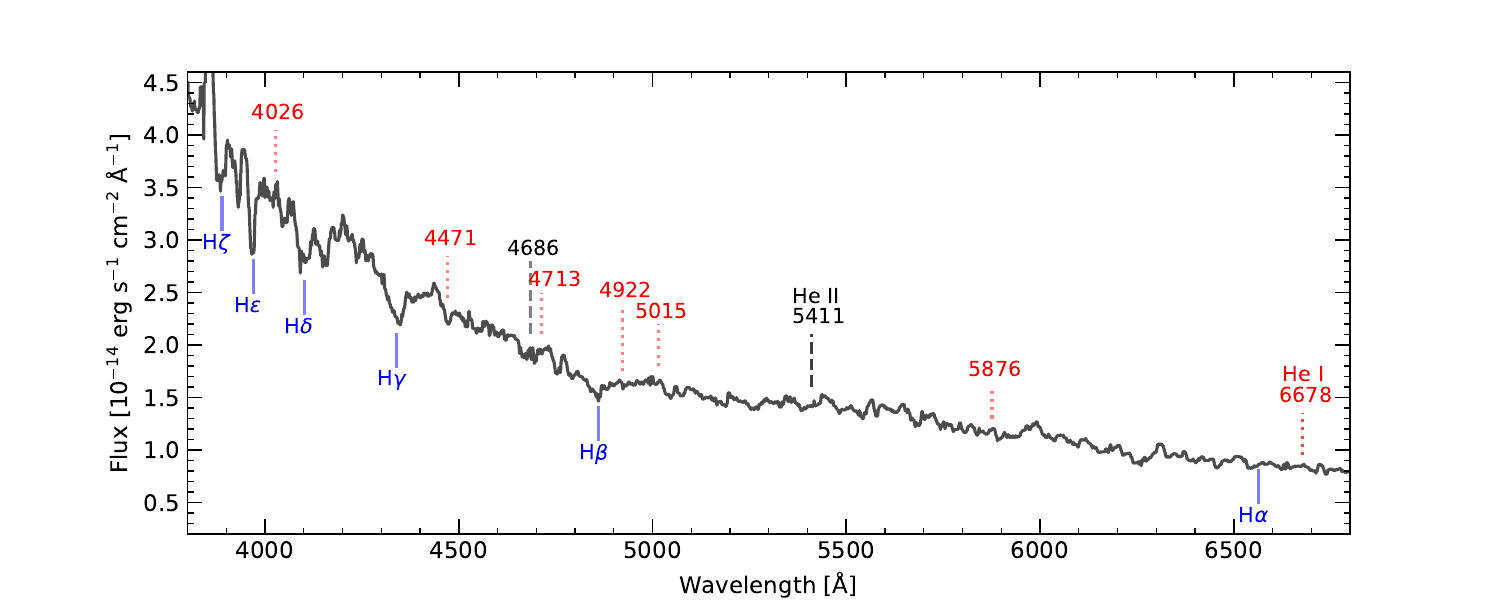}
\caption{Gemini spectrum of \target\ obtained by combining blue and red spectra
  shown after 16\AA\ smoothing.
Extinction is not corrected.
The locations of H and He lines often observed in WZ Sge-type DNe are 
  marked by vertical lines:
  solid, dotted, and dashed lines for \hi, \hei, and \heii, respectively.
They are H$\alpha$ (6563 \AA), H$\beta$ (4861 \AA), H$\gamma$ (4340 \AA), H$\delta$ (4102 \AA), 
  H$\epsilon$ (3970 \AA), H$\zeta$ (3889 \AA),
  \hei\ (red), and 
  \heii\ (black), 
  while wavelengths are shown for helium lines.
}
\label{fig:combspec}
\end{figure*}

Figure~\ref{fig:combspec} shows the combined spectrum of the blue and red spectra
  of \target\ after 16\AA\ smoothing
  obtained during its superoutburst (\S\ref{sec:discovery}).
The spectrum is featured with easily identifiable strong blue continuum
  which is dominated by the contribution from the hot and optically thick accretion disk
  \citep[e.g.,][]{Hessman84, Neustroev17, Han20}.

WZ Sge-type DNe typically exhibit somewhat complicated spectral features
  mostly from $\hi$ and $\hei$. 
While their quiescent spectra usually show
  broad Balmer series in emission often with double peaks \citep[e.g.,][]{Gilliland86,Thorstensen98,Littlefield13,Neustroev17,HernandezSantisteban19},
  spectra during superoutbursts show most of the Balmer lines in absorption largely 
  due to increased opacity with H$\alpha$ sometimes appearing as emission line
  \citep[e.g.,][]{Baba02,Nogami04,Neustroev17}. 
{\hei} lines during superoutbursts are mostly in absorption,
  whereas {\heii} 4686 {\AA} line often appears as emission \citep[e.g.,][]{Han20}.
In addition, the Bowen blend of {\hbox{C\,{\sc iii}}}/{\hbox{N\,{\sc iii}}} 
  complex at 4640 {\AA} also appears as emission in some cases \citep[e.g.,][]{Steeghs01,Nogami04}.

Figure~\ref{fig:combspec} shows strong blue continuum for \target\ together with 
  hydrogen Balmer absorption lines from H$\beta$ to H$\zeta$
  except that no identifiable feature is seen near H$\alpha$
  probably due to lower S/N ratio of our spectrum at the longer wavelength band.
This is very similar to the spectra taken at outbursts
  for a WZ Sge-type DN SSS J$122221.7-311525$ by \citet{Neustroev17}
  and for U Gem by \citet{Han20}, that show 
  strong blue continuum superposed with Balmer absorption lines
  except a weak emission in H$\alpha$.
While other {\hei} lines are not identifiable, probably due to not enough signals, 
  {\hei} $\lambda$4471 absorption line can be seen.
The spectra taken in the middle of the outburst of 2001 WZ Sge superoutburst
  shown by \citet[][their figure 4]{Nogami04} also present
  conspicuous absorption lines in H$\beta$, H$\gamma$, and H$\delta$
  together with weaker, but still noticeable, absorption lines in {\hei} $\lambda$4471
  and {\hei} $\lambda$4922.
Although very weak, {\heii} $\lambda$4686 emission line is seen in Figure~\ref{fig:combspec}
  which is often used to probe the spiral asymmetries of accretion disk \citep{Harlaftis99,Han20}.

\section{Discussion} \label{sec:discussion}   

In general, the superhump period is $\sim 1$\% longer than the orbital period
  (`positive superhumps') in WZ Sge-type DNe \citep{Osaki03, kato09_61_601, Tarasenkov25};
  however, in rare cases, superhump period has been reported to be 
  $\sim 1$\% shorter than the orbital period (`negative superhumps';
  \citealt{Smak09, Wood09, Wood11, Osaki13a, Kozhevnikov15, Sun24_974, Sun24_962,
  Pavlenko25, Vallet26}).

\begin{figure*}   
\epsscale{1.0}
\plotone{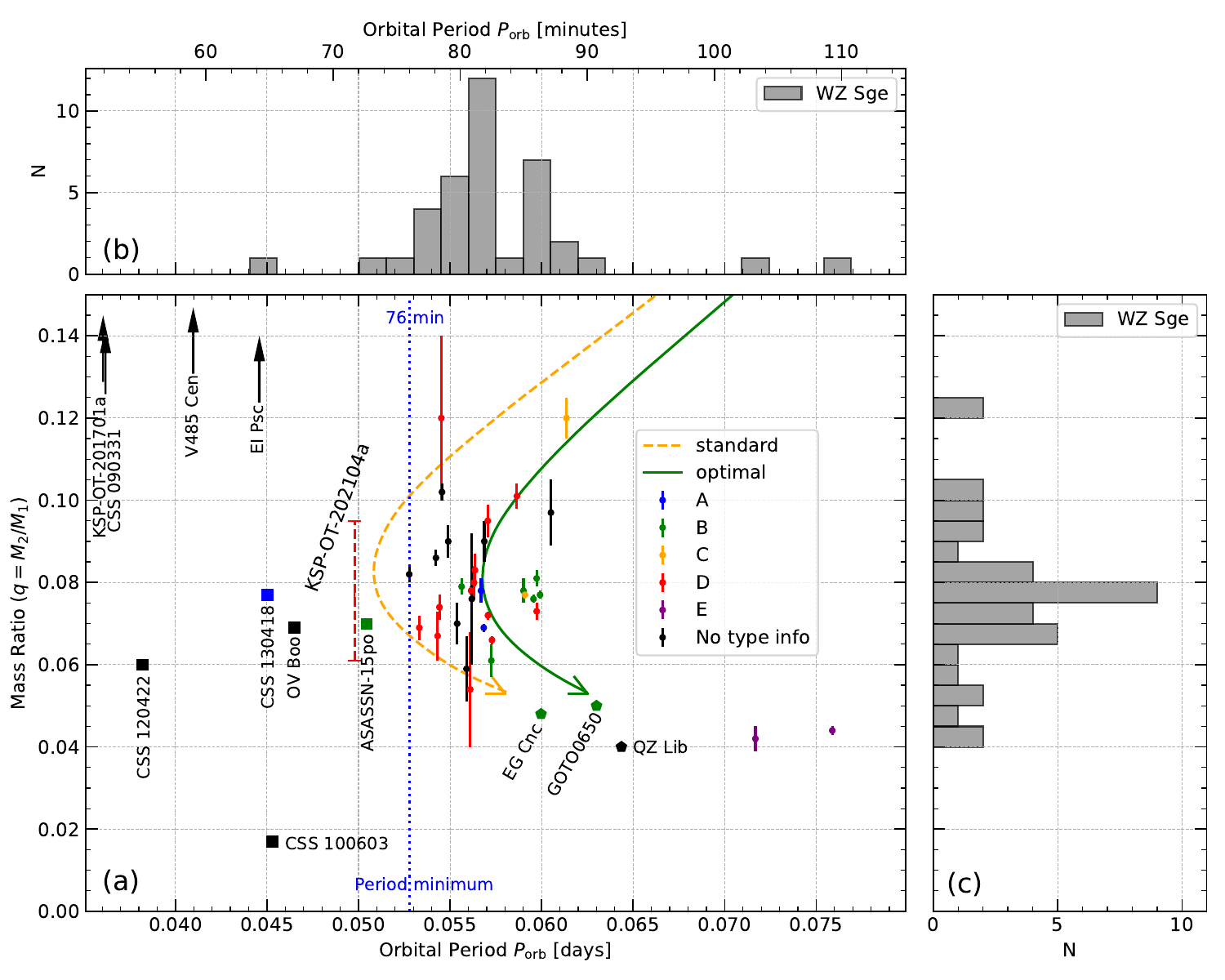}
\caption{(a) Distribution of orbital periods and mass ratios of short-period DNe:
  9 with $P_{\rm orb}$ shorter than the period minimum (large squares and black arrows), 
  34 WZ Sge-type DNe from \citet{Kato15WZSge} (dots with error bars), 
  and 3 period bouncer candidates with $P_{\rm orb}$ greater than the period minimum 
  (large pentagons).
For \target\ (red dashed line),
  we show the superhump period (the orbital period differs from the superhump period
  by $\sim$ 1\%, usually being shorter but occasionally longer) and
  the mean ($q=0.078$) and standard deviation (0.017) of the mass ratio values 
  for the known WZ Sge-type DNe from \citet{Kato15WZSge}.
Data for the first group and \target\ 
  are from Table~{\ref{tab:shortP_table}}.
The colors for the second group of WZ Sge-type DNe represent
  their types: A (blue), B (green), C (orange), D (red), 
  E (purple), and unknown type (black).
The third group of period bouncer candidates consists of 
  EG Cnc ($P_{\rm orb} = 0.060$ d, $q = 0.048$, \citealt{Kimura21}),  
  GOTO0650 ($P_{\rm orb} = 0.063$ d, $q = 0.05$, \citealt{Killestein25}), and  
  QZ Lib ($P_{\rm orb} = 0.064$ d, $q =0.040$, \citealt{Pala18}).     
The orange dashed and green solid curves show the standard and optimal evolutionary 
  tracks of DNe, respectively, from \citet{knigge11}.
The blue vertical dotted line indicates the period minimum ($\sim 76$ minutes) \citep{lee24}.
(b) Histogram of the orbital periods for the WZ Sge-type DNe.
(c) Histogram of the mass ratios for the WZ Sge-type DNe.
For panels (b) and (c), DNe known as the WZ Sge-type (CSS 130418,
  ASASSN-15po, EG Cnc, and GOTO0650) are included;
  \target\ is not. }
\label{fig:q_Porb}
\end{figure*}

Figure~\ref{fig:q_Porb} shows the distribution of the orbital periods and 
  mass ratios of several WZ Sge-type DNe from \citet{Kato15WZSge} and other DNe with short periods
  compared with theoretical evolutionary tracks from \citet{knigge11}:
  orange dashed and green solid curves for the standard and optimal tracks, respectively.
Here, the `optimal' evolutionary tracks are the improved version
  of the standard model modified to match to the data, especially below
  the period gap \citep{knigge11}. 
As delineated by the evolutionary tracks, 
  the orbital periods decrease down to the period minimum of $\sim$ 76 minutes
  as a result of angular momentum loss of the secondaries
  caused by continuous mass transfer.
The mass ratio also decreases along these processes.
Near the period minimum, the increasing electron degeneracy in the secondary
  together with its decreasing density lead to the transition from 
  a main-sequence star to a brown dwarf \citep{Kolb99}. 
During this transition, the orbital period starts to increase, 
  known as the ``period bounce'' \citep{Kolb99,Pala20}.
As shown in Figure~\ref{fig:q_Porb}, the DNe EG Cnc \citep{Kimura21},
  GOTO0650 \citep{Killestein25}, and QZ Lib \citep{Pala18}  
  are candidate ``period bouncers'' that recently passed the period bounce 
  \citep{Kato13,Patterson98,Pala20,Neustroev23}.

In Figure~\ref{fig:q_Porb}, since the orbital period and the mass ratio
  are not known for \target, 
  we instead used the superhump period (usually $\sim 1$\% longer than the orbital period)
  and the mean and standard deviation values of the mass ratios
  known for the WZ Sge-type DNe from \citet{Kato15WZSge}.
CVs with orbital periods shorter than the period minimum,
  possibly including the DN like \target, are believed
  to be composed of a WD and a small-mass secondary star \citep{Green20,lee22}.
It is generally believed that these small-mass secondaries have 
  evolved to high-density, He-rich stars via significant mass transfers
  by which their atmospheres are stripped away \citep[e.g.,][]{Green20,Green25}.
The most common samples for these He-rich secondaries are He CVs and AM CVn stars
  exhibiting He-rich spectra with H abundance being much smaller
  in the latter than the former \citep{Green20}.
AM CVn stars are considered to be composed of a WD primary and a secondary star,
  which is (1) a less massive WD, (2) a low-mass semi-degenerate He-burning star,
  or (3) an evolved main-sequence star with its H-envelope stripped away
  \citep{Solheim10,Breedt12,Green20}.

While we denoted the period minimum to be around 76 minutes 
  \citep{Knigge06, Kato_etal15} in Figure~\ref{fig:q_Porb},
  this value is not accurately known yet.
Observations usually report $\sim 80-82$ min values for it
  \citep{Gansicke09}, 
  while it is theoretically predicted to be around $65-70$ min
  \citep{Kolb99, Howell01, knigge11}.
Further efforts are needed to resolve this `period minimum problem',
  or to clarify whether the period minimum intrinsically spans a range
  \citep{Kato_etal15}.  

Table~\ref{tab:shortP_table} lists the parameters of 
  ten DNe (including \target) with orbital periods shorter than the period minimum (76 minutes)
  that we have collected from the literature search.
As shown in Figure~\ref{fig:q_Porb}, the nine DNe with orbital periods
  shorter than the period minimum
  show a broad range in the mass ratio, ranging from $q$ $\simlt$ 0.02 (CSS 100603)
  to $q$ $>$ 0.10 (KSP-OT-201701a, CSS 090331, V485 Cen, and EI Psc).
Although these objects, which often show H-poor spectra,
  are largely considered to be evolving to 
  AM CVn or He CV stars \citep{Solheim10,Littlefield13,Green18mn,lee22,lee24}
  from long-period DNe, 
  their exact evolutionary nature remain to be understood.

\begin{deluxetable*}{ccccc}\label{tab:shortP_table}
	\tablecaption{Parameters of DNe with orbital period ($P_{\rm orb}$) shorter than the period minimum}
	\tablehead{\colhead{Name} & \colhead{Other Name} & 
                        \colhead{$P_{\rm orb}$ (min)} & \colhead{$q=M_2/M_1$} & 
		\colhead{Reference}
	}
\startdata
ASASSN-15po    &                 & 72.6  & 0.0699 & \citet{Namekata17} \\
\target\ && $71.7$\tablenotemark{a} & --- & This study \\
OV Boo            & SDSS 1507+5230 & 66.6 & 0.069 & \citet{Littlefair07}, \\
&&&& \citet{Patterson08}, \citet{Uthas11} \\
CSS 100603       & CSS 1122$-$1110  & $65.233\pm0.015$ & $0.017\pm0.004$ & \citet{Breedt12} \\
CSS 130418  & CSS 174033.5+414756 & $64.84\pm0.01$  & $0.077\pm0.005$ & \citet{Chochol15}, \citet{Imada18} \\
CSS 120422       & CSS 111127+571239, SBS 1108+574 & 55  & 0.06 & \citet{Littlefield13} \\
\hline 
EI Psc              & RX 2329, 1RXS J232953.9+062814 & 64.2 & 0.185 & \citet{Thorstensen02} \\
V485 Cen          &                            & 59.0     & 0.38 & \citet{Augusteijn93,Augusteijn96} \\
CSS 090331       & CRTS J1028$-$0819  & $52.1\pm0.6$ & $0.25\pm0.06$ & \citet{kato09Dec}, \\
&&&& \citet{Woudt12}, \citet{Green20} \\
KSP-OT-201701a &              & $51.91 \pm 2.50$ & 0.37 & \citet{lee22} \\
\enddata
\tablenotetext{a}{Superhump period.}
\end{deluxetable*}

\section{Summary and Conclusion} \label{sec:summary}  

In this work, we present the discovery and photometric and spectroscopic studies of \target,
  which is a new WZ Sge-type DN, 
  possibly under the period minimum from KSP during its superoutburst.
We summarize our results as below. 
\begin{itemize}
\item 
  The peak $V$-band magnitude of the superoutburst is $\sim$ 16.01 mag
  which is about 8 magnitudes brighter than its quiescent brightness. 
  The duration of the superoutburst is $\sim 28.5$ days featured with a plateau
  without rebrightening, identifying it to be a type D  WZ Sge-type DN.
We obtain its superhump period to be $P_{\rm sh} \approx 71.7$ minutes ($=0.04978$ days).
\item
{\target} might be a source with orbital period 
  shorter than the ``period minimum'' of $\sim$ 76 min,
  considering the fact that the superhump and orbital periods have a very similar value
  ($\sim$ 1\% difference).
The small number of objects in this period region are 
  transitional ultracompact accreting binaries \citep{Green20}
  that may be in the evolution between longer period DNe (with main-sequence companion stars)
  and AM CVn stars with very short orbital periods.
If the orbital period of {\target} is proved to be shorter than the period minimum, 
  this DN with a small mass transfer may also be evolving toward AM CVn.
\item 
Our spectrum of the source clearly shows blue continuum 
  from the optically thick accretion disk during the superoutburst
  with hydrogen absorption lines from H$\beta$ to H$\zeta$,
  while signal at the wavelength near H$\alpha$ is too weak to identify the nature.
\end{itemize}


\begin{acknowledgments}
We deeply appreciate the anonymous referee
  for the suggestions and detailed comments
  that greatly helped improve the manuscript.
This research has made use of the KMTNet system 
  operated by the Korea Astronomy and Space Science Institute (KASI) 
  at three host sites of CTIO in Chile, SAAO in South Africa, and SSO in Australia.
Data transfer from the host site to KASI was supported by
  the Korea Research Environment Open NETwork (KREONET).
The Gemini South observations were obtained 
  under the K-GMT Science Program (PID: GS-2021A-Q-117, PI : Hong Soo PARK) of KASI. 
This research was supported by the Korea Astronomy and Space Science Institute under the R\&D program 
  (Project No. 2026-1-860-00) supervised by the Ministry of Science and ICT.
Y.L. was supported by Basic Science Research Program through the National Research Foundation
  of Korea (NRF) funded by the Ministry of Education (NRF-2022R1I1A1A01054555).
D.-S.M. was supported in part by a Leading Edge Fund from the Canadian Foundation
for Innovation (project No. 30951) and a Discovery Grant (RGPIN-2019-06524) from the Natural Sciences
and Engineering Research Council (NSERC) of Canada.
\end{acknowledgments}


\bibliography{draft}
\bibliographystyle{aasjournalv7}

\end{document}